 \documentstyle[aas2pp4]{article}
\input psfig.tex
\def \m{\ifmmode M_\odot\else M$_\odot$\fi}
\def \r{\ifmmode R_\odot\else R$_\odot$\fi}

\def \gta {\mathrel{\vcenter
     {\hbox{$>$}\nointerlineskip\hbox{$\sim$}}}}

\def\beq{\begin{equation}}
\def\eeq{\end{equation}}
\def\ref{\reference}
\def\gr{$\gamma$-ray }
\def\grb{$\gamma$-ray burst }

\begin{document}
\title{SN 1998bw/GRB 980425:} 
\title{Hypernova or Aspherical Explosion?}
\author{ Peter H\"{o}flich,  J. Craig Wheeler,  Lifan Wang}
\affil{Department of Astronomy, University of Texas, Austin, TX 78712, USA}   
\affil{E-Mail: pah@alla.as.utexas.edu, lifan@tao.as.utexas.edu, wheel@alla.as.utexas.edu}

\begin{abstract}
 The recent discovery of the unusual supernova SN1998bw and its apparent 
correlation with the gamma-ray burst GRB 980425 has raised new issues 
concerning both the GRB and SNe. 
SN1998bw was unusually bright  at maximum light and
expansion velocities were large
 making SN1998bw a 
possible candidate for a ``hypernova" with explosion energies exceeding $10^{52} erg$.
We show that the light curve  of SN1998bw
can be understood as the result of an aspherical explosion along the rotational 
axis of a basically spherical, non-degenerate C/O core of a massive star with
an explosion energy of $2 \times 10^{51}$ erg,  a total ejecta mass of $2 M_o$, and a $^{56}Ni$ mass
of $0.2M_\odot$ if it is seen from 
high inclinations with respect to the plane of symmetry. In this model, the high
expansion velocities are a direct consequence of the  aspherical explosion which, 
in turn, produces oblate iso-density  contours. This suggests that the fundamental 
core-collapse explosion process itself is strongly asymmetric.

\end{abstract} 
 
\keywords{Supernovae: general, individual (SN~1998bw) --- 
gamma-ray bursters --- radiation transfer --- asphericity }

\section{Introduction}
 
Due to its correlation in time and location, the $\gamma $-ray burst GRB~980425
has a high probability of being associated with SN~1998bw (Galama et al. 1998).
This connection is supported by the association of
a relativistically expanding radio source with
SN~1998bw (Kulkarni et al. 1998). 
From optical spectra, SN~1998bw was classified as a 
SN~Ic by Patat and Piemonte (1998).  
What sets SN~1998bw apart from other SNe~Ic 
are  higher expansion 
velocities as indicated by the Si II and Ca H and K lines
($\approx $30 to 50 percent higher at maximum light than  SN1994I and SN1983V,
 Clochiatti  \& Wheeler 1997), the red colors at maximum light, and the 
large intrinsic brightness.
A peak luminosity of $1.3\pm 0.6~10^{43} erg/s$  can be inferred from the redshift 
of the host galaxy ($z=0.0085$, Tinney et al. 1998) and the reddening
($A_V=0.2^m$, Schlegel et al. 1998), if we assume $H_o = $ 67 km/s/Mpc.
 The uncertainties are rather large as the host galaxy is not
 yet fully in the Hubble flow, so  the peculiar velocity may 
be of the order of 300 to 400 km/sec,  $H_o$ is known
only to an accuracy of $\approx 10 \%$, and 
 $A_V$  may vary by $\approx 0.1^m$.
 
The properties of 
 SN~1998bw  suggest that it was a ``hypernova" event 
(Paczy\'nski 1997). Based on their light curve (LC) calculations,
Iwamoto et al. (1998) and Woosley et al. (1998) derived explosion energies of
20-50 foe and 22foe ($1 foe = 10^{51} erg$)
 ejecta masses of 12-15 $M_\odot$ and 6 $M_\odot$, 
and $^{56}$Ni masses of 0.6-0.8 and 0.5 $M_\odot$, respectively.
 These models have some problems. Although Iwamoto's fit of the 
LC is excellent with errors $\leq 0.3^m $ over 40 days, the spectra
show absorption lines that are too narrow by a factor of 2 to 3 indicating  too narrow a
range of formation in velocity space. This may be related to the high envelope mass.
In the lower mass models of Woosley  et al. (1998),
 the bolometric and monochromatic LCs 
differ from the observations by 0.5 to 1 magnitude over the course  
of 20 days and all the computed color indices (B-V, V-R, V-I)
are too red by about the same amount at all epochs.
 
Guided by the deduced properties of more traditional core 
collapse SNe and SNe~Ic  in particular, 
we want to demonstrate that asphericity is  an alternative explanation 
for SN~1998bw that puts it back into the range of ``normal" SNe~Ic.
 
 Both spectral analyses and 
LC calculations support the picture that
SNe~II, SNe~Ib and SNe~Ic may form a sequence involving core 
collapse with successively smaller H and He envelopes
(Clochatti \& Wheeler 1997).
The analysis of spectra and LCs gives essentially no
insight into the geometry of the expanding envelope. 
Polarization, however, provides a unique tool to explore asymmetries.
Linear polarization of $\approx 1 \% $ seems to be typical for SNe~II
 (Wang, Wheeler \& H\"oflich
 1998).  There is a trend, however, for the observed polarization 
to increase in core-collapse SNe with decreasing envelope mass,
e.g. from SN~II to SN~Ic (Wang et al. 1998).  
For SN~1993J,   the observed linear polarization 
was as high as $\approx 1.0 ... 1.5 \% $, and 
for the SN~Ic 1997X the polarization was even higher 
(Wang et al. 1998). 
This trend, while tentative, clearly points toward the 
interpretation that the explosion itself is strongly asymmetric. 
 
From theoretical calculations for scattering dominated atmospheres, 
this size of polarization, $\gta 1 $ \%,
requires axis ratios of the order of 2 to 3, 
making these objects highly aspherical. The  luminosity 
$L(\Theta )$ 
 will change by a factor of
 $\approx $ 2 as the line of sight varies from the equator to the pole 
(H\"oflich 1991, H\"oflich et al. 1995).  
 
Given the ubiquitous presence of polarization in core collapse
supernovae and especially SN~Ic,  inclusion of asphericity effects in SN~Ib/c may 
prove to be critical (Wang et al. 1998). Polarization
has been observed in SN1998bw (Kay et al. 1998).
 
In this work, we present a first approach to the problem of asymmetric 
LCs for SN~Ic and SN~1998bw.

\section{Description of the Concept and  Numerical Methods}

For the initial setup, we use the chemical and density 
structures of spherical C/O cores 
 of Nomoto and Hashimoto (1988). 
These structures are scaled to
adjust the total mass of the ejecta.
This is an approximation, but the details 
of the chemical profiles are not expected to effect the light 
curves. 
 
The explosion models are calculated using a one-dimensional
radiation-hydro code which 
includes  a detailed nuclear network.
The code also simultaneously solves for the radiation transport via moment equations.
  Photon redistribution and
thermalization is based on detailed NLTE-models. Several hundred frequency 
groups are used to calculate  monochromatic LCs, frequency-averaged Eddington factors
and opacity means.
A Monte Carlo scheme is used for $\gamma$-rays.
For details, see H\"oflich, Wheeler \& Thielemann 
 (1998) and references therein.

 Aspherical density structures are constructed based on the 
spherical density distribution.  For the simple models presented
here,  we impose the asymmetry after the ejecta has
reached the homologous expansion phase.  We generate an
asymmetric configuration by preserving the mass fraction
per steradian from the spherical model, but imposing a different
law of homologous expansion as a function of the angle $\Theta$
from the equatorial plane.
For typical density structures, a higher energy deposition 
along the polar axis results 
in oblate density structures.
Such an energy pattern may be produced if jet-like structures 
are formed during the central core collapse as suggested by 
Wang \& Wheeler (1998). In contrast, a prolate density structure
would be produced if more energy is released in the equatorial region
than in the polar direction. For more details, see H\"oflich, Wang \& Wheeler (1998).

The bolometric and broad band LCs are constructed 
by convolving the spherical LCs with the photon redistribution functions L$(\Theta)$/L(mean)
 which are calculated
by our Monte Carlo code for polarization (H\"oflich 1991, H\"oflich et al. 1995).
Typical conditions at the photosphere and therefore the colors 
are expected to be similar in both the spherical and aspherical models
because the energy flux $F(\Theta)$ is found to be similar both 
in the spherical and aspherical configuration to within $\approx 40 \% $. 
To first order (Wien's limit), a change of $F(\Theta)$  by 40 \% 
corresponds to a change in color indices  by about 0.1$^m$.
Stationarity is assumed to calculate the photon redistribution functions
since the geometry does not change during the typical diffusion 
time scale.
We assume implicitly the same
mean diffusion time scales for both the spherical and aspherical configurations.
This mostly effects the very early phases of the LC when the
hydrodynamical time scales are short. For more details, see H\"oflich, Wang \& Wheeler
(1998).

\section{Results}
 
We construct models in such a way a way that at day 20
the axis ratio at the photosphere is 2.
In comparison to the spherical model, the homology expansion parameters are 
a factor of $\approx 2.2 $ larger along the pole for oblate ellipsoids 
and a factor of $\approx 1.5$ larger for prolate ellipsoids 
along the equator.
 
We first calculated aspherical LCs based 
on the C/O core CO21  which gives a good representation of the BVRI LCs of the SN~Ic 1994I
(Iwamoto et al. 1996). The ejecta mass is $0.8 M_\odot $ and the explosion energy is
$E_{kin} = 10^{51} erg $. A mass of
$0.08~ M_\odot $ of $^{56}Ni$ is ejected.
 This model failed  to produce  the
peak brightness by a factor of 2, gave too short a rise time by about 5 days, and shows
blue color indices at maximum light.

To boost the total luminosity to the level of observation 
we increased the amount of ejected  $^{56}Ni$ to 0.2 $M_\odot$. 
This quantity of nickel is still below the estimates for 
$^{56}Ni$ of 0.3 $M_\odot$ in the bright SN~II 1992am (Schmidt et al. 1997). 
As shown in Fig.1,  the time of maximum light is 
rather insensitive to asphericity effects.
The need to delay the time to maximum and to produce  the red color at maximum light
suggested the need to increase the ejecta mass with
an appropriate increase in the kinetic energy to provide the
observed expansion.
We thus computed a series of models with $M_{ej}=2 M_\odot$ 
$E_{kin}=2\times10^{51}$ erg and $M_{Ni} = 0.2$\m. 
 \begin{figure}[t]
  \psfig{figure=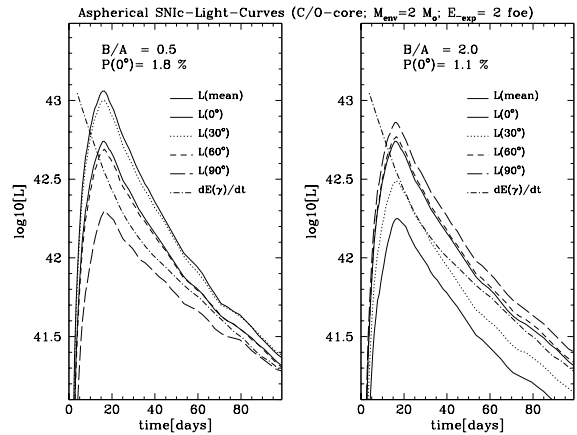,width=8.0cm,rwidth=7.0cm,clip=,angle=270}
 \caption{Directional dependence of the bolometric LC for oblate (left) and prolate
 (right) ellipsoids. The luminosity of the corresponding spherical model is shown as $L(mean)$.
 In addition, the instantaneous $\gamma-ray$ deposition is shown.
 $P(0^o)$ is the polarization at maximum light ($ P(\Theta) \approx P(0^o) \times 
 cos^2\Theta$).}
  \end{figure}
 
Asphericity of the amplitude assumed here
can change the luminosity over a range of roughly 2 magnitudes (Fig. 1).
For oblate ellipsoids, the luminosity is
enhanced along the pole whereas for prolate structures 
the enhancement occurs in the equatorial direction. 
Combined with the polarization properties, 
this provides a clear separation between oblate and prolate geometries as 
 $P$ always goes to 0  if the structure is seen pole-on and  $P$  
increases towards lower latitudes (H\"oflich 1991). 

 Observations of the polarization of SN~1998bw 23 days after
the explosion show little polarization ($< 1. \% $, Patat et al. 1998). 
By day 58,
the intrinsic polarization was reported to be 0.5 percent (Kay et al. 1998).  
Polarization data on SN~Ic is rare, but this value is less than seen in some SNe~Ic
and related events
(see above). SN~1998bw was also rather bright.
This combination implies oblate geometries if asymmetry is involved.

For the comparison between the observed and theoretical LCs (Fig. 2),
we have used the relative calibration of Woosley et al. (1998)
for the ''bolometric LC".  The broad-band data was obtained from 
Galama et al (1998).  Apparently, the object
must be seen from an angle of $\ge 60^o$ from the equator. 
The same conclusion can be drawn independently from 
the  detected but relatively small linear polarization.
 
 Overall, the broad-band LCs agree with the data within the 
uncertainties.
The intrinsic color excess B-V matches the observations within $0.1^m$ 
and, after the initial rise of $\approx 7$ days, the agreement in each band
 is better than $0.3^m$. 
The main discrepancy with the observations occurs during 
the initial rise when the diffusion time scales are
much longer than the expansion time.  Under these conditions, 
our approximation of redistribution of the energy
of a spherical model breaks down since the diffusion time scale 
is long compared to the hydrodynamical time scale.
The decline after maximum is slightly too steep in the models both in the 
bolometric and broad band LCs.
This is likely to be related to the energy generation in the 
envelope by \gr ~deposition or to the change in the escape 
probability of low energy photons. 
 The decline rate immediately after peak 
can be reduced by increasing the amount of radioactive $^{56}Ni$ by
 $\approx 40 \%$. 
In our models, the escape probability for $\gamma $-rays 
 increased rapidly between day 20 and 80 from 3\% to 50 \%.
 An alternative means to flatten the light curve is to reduce the increase in 
 escape probability. This can be achieved by a modification restricted to the inner
layers of the ejecta because the escape probability 
 is determined by those layers.
Either the expansion velocity of the inner layers can be reduced or the density
gradient  may become steeper. Both are expected for strongly aspherical explosions.
  \begin{figure}[t]
   \psfig{figure=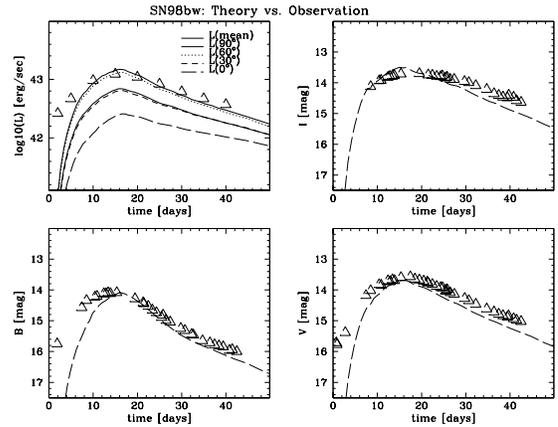,width=8.0cm,rwidth=7.0cm,clip=,angle=270}
  \caption{Comparison of bolometric and broad-band  LCs as observed for 
  SN1998bw with those of an oblate ellipsoid seen at high angle with respect to the equator
 assuming a distance of 36Mpc and  $A_V=0.2^m$.}
  \end{figure}

\section{Discussion and Conclusions}
 
We have shown that the high apparent luminosity of SN~1998bw 
may be understood within the framework of ``classical" SNIc. 
Even with our current model, SN~1998bw remains 
at the bright end of the scale.  We note that the luminosity of 
SN~1998bw may be uncertain by a factor of 2 due to non-Hubble
motion within the cluster and uncertainties in the Hubble constant and reddening. 
For a model with an ejected mass
of 2 $M_\odot$, an explosion energy of $2\times 10^{51}$ erg, and 
a $^{56}Ni$- ejection of $0.2 M_\odot$,
both the bolometric and broad-band LCs 
are rather well reproduced by an oblate 
ellipsoid with an axis ratio of 0.5 which is observed within 
$30\deg$ of the symmetry axis.
This angle for the line of sight is consistent with the low 
(but still significant) polarization observed for SN~1998bw. 
In a Lagrangian frame, the polar expansion velocity is a factor of 2 larger
than the mean velocity. This is also in agreement with the rather 
large expansion velocities seen in SN~1998bw.
 
Woosley et al. (1998) have analyzed the possibility of 
$\gamma $-ray bursts in the framework
of spherical models. Even with their explosion energies of more 
than 20 foe they showed that the \grb\ associated with
SN~1998bw/GRB~980425 cannot be explained by the acceleration
of matter to relativistic speeds at shock-breakout. 
In our picture, the specific energy released in the polar region 
is comparable.
 We want to stress, however, that our asymmetry in the energy distribution is
set after homologeous expansion is established. 
Because the early hydrodynamical evolution will rather
tend to wipe out asymmetries, 
the initial anisotropy in the energy distribution
is expected to be significantly higher. 
How much, only detailed multi-dimensional hydro-calculations can tell.
 
We have shown that SN~1998bw may be 
understood within the framework of ``classical" core collapse supernovae 
rather than by a ,``hypernova", but the actual model parameters must be regarded as uncertain
both because of the model assumptions and the uncertainty in the observed luminosity.
 In light of the good fits of 
Iwamoto et al. (1998), however, SN~1998bw may indeed be a ,''hypernova".
Continuous measurements
of the polarization and the velocity of $^{56}Co$ lines
are critical to unreveal the nature and geometry of this object.
 For more details, see H\"oflich, Wheeler \&  Wang (1998).

\subsection*{ACKNOWLEDGMENTS} 

We thank  Ken Nomoto for providing us with the monochromatic LC data in 
digital form. This research was supported in part by NSF Grant AST 9528110, 
NASA Grant NAG 5-2888, and a grant from the Texas Advanced Research Program.

\end{document}